\documentclass[aoas,nameyear,dvips]{arximspdf}
\usepackage{dcolumn,multirow}
\usepackage{graphicx}

\doi{10.1214/10-AOAS357}
\volume{4}
\issue{4}
\pubyear{2010}
\firstpage{1749}
\lastpage{1773}

\makeatletter
\newcommand{\sign}{\operatorname{sign}}
\newcolumntype{d}[1]{D{.}{.}{#1}}
\newcolumntype{b}[1]{D{.}{\ }{#1}}
\newcolumntype{f}[1]{D{.}{\,}{#1}}
\newcolumntype{k}[1]{D{.}{}{#1}}
\makeatother

\begin{document}
\begin{frontmatter}

\title{Reconstructing DNA copy number by penalized estimation and imputation}
\runtitle{CNV by penalized estimation}

\begin{aug}
\author[A]{\fnms{Zhongyang} \snm{Zhang}\thanksref{t1}\ead[label=e1]{zhangzy@ucla.edu}},
\author[B]{\fnms{Kenneth} \snm{Lange}\thanksref{t2}\ead[label=e2]{klange@ucla.edu}},
\author[C]{\fnms{Roel} \snm{Ophoff}\thanksref{t1}\ead[label=e3]{ophoff@ucla.edu}}\\
\and
\author[D]{\fnms{Chiara} \snm{Sabatti}\corref{}\thanksref{t1,t2}\ead[label=e4]{sabatti@stanford.edu}}
\thankstext{t1}{Supported in part by NIMH Grant MH078075-01.}
\thankstext{t2}{Supported by NIH/NIGMS Grant GM053275-14.}
\runauthor{Zhang, Lange, Ophoff and Sabatti}
\affiliation{University of California, Los Angeles}
\address[A]{Z. Zhang\\
Department of Statistics\\
University of California, Los Angeles\\
Los Angeles, California 90095\\
USA\\
\printead{e1}} 
\address[B]{K. Lange\\
Department of Biomathematics, \\
\quad Human Genetics and Statistics\\
University of California, Los Angeles\\
Los Angeles, California 90095\\
USA\\
\printead{e2}}
\address[C]{R. Ophoff\\
Center for Neurobehavioral Genetics\\
University of California, Los Angeles\\
Los Angeles, California 90095\\
USA\\
\printead{e3}}
\address[D]{C. Sabatti\\
Departments of HRP and Statistics\\
Stanford University \\
California 94305\\
USA\\
and\\
Departments of Human Genetics \\
\quad and Statistics\\
University of California, Los Angeles\\
Los Angeles, California 90095\\
USA\\
\printead{e4}}
\end{aug}

\received{\smonth{6} \syear{2009}}
\revised{\smonth{3} \syear{2010}}

\begin{abstract}
Recent advances in genomics have underscored the surprising ubiquity of
DNA copy number variation (CNV). Fortunately, modern  genotyping
platforms also detect CNVs with fairly high reliability. Hidden Markov
models and algorithms have played a dominant role in the interpretation
of CNV data. Here we explore CNV reconstruction via estimation with a
fused-lasso penalty as suggested by Tibshirani and Wang [Biostatistics \textbf{9} (\citeyear{cghFLasso}) 18--29]. We mount a fresh
attack on this difficult optimization problem by the following: (a)
changing the penalty terms slightly by substituting a smooth
approximation to the absolute value function, (b) designing and
implementing a new MM (majorization--minimization) algorithm, and (c)
applying a fast version of Newton's method to jointly update all model
parameters. Together these changes enable us to minimize the
fused-lasso criterion in a highly effective way.

We also reframe the reconstruction problem in terms of imputation via
discrete optimization. This approach is easier and more accurate than
parameter estimation because it relies on the fact that only a handful
of possible copy number states exist at each SNP. The dynamic
programming framework has the added bonus of exploiting information
that the current fused-lasso approach ignores. The accuracy of our
imputations is comparable to that of hidden Markov models at a
substantially lower computational cost.
\end{abstract}

\begin{keyword}
\kwd{$\ell_1$ penalty}
\kwd{fused lasso}
\kwd{dynamic programming}
\kwd{MM algorithm}.
\end{keyword}

\end{frontmatter}
\section{Introduction}

New techniques of fine mapping have uncovered many regions of the human
genome displaying copy number variants (CNVs)
[\citet{Iafrate04}; \citet{redon06}; \citet{sebat04}]. Variation is to be expected in
cancer cells, but it also occurs in normal somatic cells subject to
Mendelian inheritance. As awareness of the disease implications of CNVs
has spread, geneticists have become more interested in screening their
association study samples for copy number polymorphisms (CNPs)
[\citet{nature}]. Fortunately, the Illumina and the Affymetrix platforms
used in high-density genotyping yield CNV information at no additional
cost. Despite their obvious technical differences, the two platforms
generate conceptually very similar CNV reconstruction problems.

Hidden Markov models and algorithms have dominated the field of CNV
reconstruction [\citet{colella07}; \citet{bird}; \citet{ingo}; Wang et al. (\citeyear{pennCNV}, \citeyear{us})]. This
statistical framework is flexible enough to accommodate several
complications, including variable single nucleotide polymorphism (SNP)
frequencies, variable distances between adjacent SNPs, linkage
disequilibrium and relationships between study subjects. In the current
paper we  investigate the potential of penalized estimation for CNV
reconstruction.  \citet{cghFLasso} introduced the fused-lasso penalty
for the detection of CNVs based on generic considerations of smoothness
and sparsity [\citet{rof}; \citet{FLasso}]. The application of the fused lasso to
CNV detection is best motivated by a simplified model. Let the
parameter vector $\bolds{\beta}=(\beta_1,\beta_2, \ldots,\beta_n)$
quantify DNA levels at $n$ successive SNPs. These levels are normalized
so that $\beta_i=0$ corresponds to the standard copy number 2, where
SNP $i$ is represented once each on the maternal and paternal
chromosomes. Variant regions are rare in the genome and typically
involve multiple adjacent SNPs; CNVs range from a few thousand to
several million base pairs in length. In high-density genotyping we
query SNPs that are on average about five thousand base pairs apart.
The true $\bolds{\beta}$ is therefore expected to be piecewise
constant, with the majority of values equal to $0$ and a few segments
with positive values (indicating duplication) and negative values
(indicating deletion).

\citet{cghFLasso} proposed the joint use of a lasso and a fused-lasso
penalty $p(\bolds{\beta})= \sum_{i=2}^{n}|\beta_i-\beta_{i-1}|$ to
enforce this piecewise constant structure. One then estimates
$\bolds{\beta}$ by minimizing the objective function
$l(\bolds{\beta})+\lambda_1 \|  \bolds{\beta}\|_{\ell_1}
+\lambda_2 p(\bolds{\beta})$, where $l(\bolds{\beta})$ is a
goodness-of-fit criteria. The nondifferentiability of the objective
function makes minimization challenging [\citet{pco}]. We mount a fresh
attack on this difficult optimization problem by the following tactics:
(a)~changing penalty terms slightly by substituting a smooth
approximation to the absolute value function, (b) majorizing the
substitute penalties by quadratics and implementing a new MM
(majorization--minimization) algorithm based on these substitutions, and
(c) solving the minimization step of the MM algorithm by a fast version
of Newton's method. When the loss function is quadratic, Newton's
method takes a single step. More radically, we also reframe the
reconstruction problem in terms of imputation via discrete
optimization. Readers familiar with Viterbi's algorithm from hidden
Markov models will immediately recognize the value of dynamic
programming in this context.  For the specific problem of detection of
CNVs in DNA from normal cells, discrete imputation has the advantage of
choosing among a handful of copy number states rather than estimating a
continuous parameter.  This fact renders discrete imputation easier to
implement and more accurate than imputation via parameter estimation.

The remainder of the paper is organized as follows. In the methods
section we briefly review the data generating mechanism for CNV
problems. We then present our estimation approach to CNV reconstruction
and the MM algorithm that implements it. Finally, we describe our new
model and the dynamic programming algorithm for discrete imputation. In
the results section we assess the statistical performance and
computational speed of the proposed methods on simulated and real data
sets.

\section{Methods}

\subsection{Characteristics of the genotype data}\label{data}

When reconstructing CNV from genotype data, researchers rely not only
on the final genotype calls but also on raw measurements obtained from
the genotyping array. The character of these measurements varies
slightly depending on the platform adopted. For definiteness, we focus
on the data delivered by the Illumina platform at our disposal. A DNA
sample from an individual is preprocessed, hybridized to a chip, and
queried at $n$ SNPs. For convenience, we will call the two alleles A
and B at each SNP. The amount of DNA carried by each allele at a
queried SNP is measured by recording the luminescence of specifically
labeled hybridized DNA fragments. Transformations and normalizations of
the luminescences lead to two noisy measurements for each SNP $i$:
$y_i$ (LogR) and $x_i$ (BAF). The former quantifies the total DNA
present at the SNP. After normalization, the average of $y_i$ across
individuals is 0.  A large positive value suggests a duplication; a
large negative value suggests a deletion. The variability $y_i$ has
been successfully described as a mixture of a Gaussian  and a
distribution to guard against contamination from outliers
[\citet{colella07}; Wang et al. (\citeyear{pennCNV}, \citeyear{us})].

\begin{figure}

\includegraphics{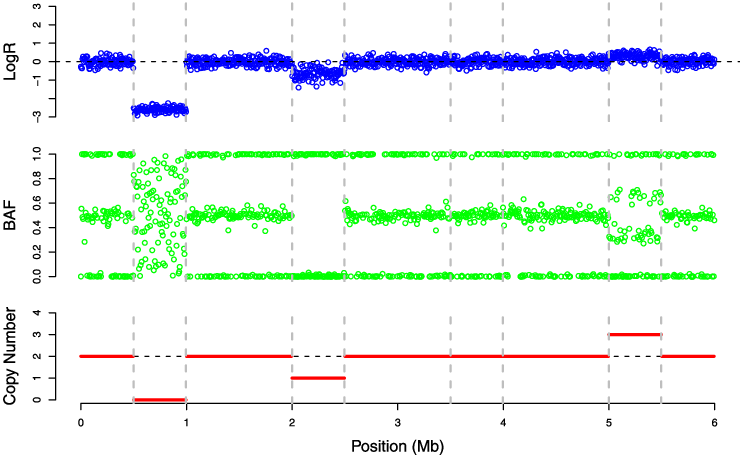}

\caption{Signal patterns for different DNA copy number scenarios
organized by their physical locations along a simulated chromosome. The
top panel displays in blue $y_i$ (LogR), the middle panel displays in
green $x_i$ (BAF), and the bottom panel displays in red the true copy
number.} \label{fig1}
\end{figure}

The B-allele frequency (BAF) represents the fraction of the total DNA
attributable to allele B. The admissible values for $x_i$ occur on the
interval $[0,1]$. When copy number equals 1, $x_i$ takes on values
close to 0 or 1, corresponding to the genotypes A and B. When copy
number equals 2, $x_i$ is expected to fluctuate around the three
possible values 0, 1/2 and 1, corresponding to the three possible
genotypes AA, AB and BB.  When copy number equals 3, $x_i$ varies
around the four possible values 0, 1/3, 2/3, 1, corresponding to the
genotypes AAA, AAB, ABB, BBB.  When copy number equals 0, the value of
$x_i$ is entirely due to noise and appears to be distributed uniformly
on $[0,1]$. Figure \ref{fig1} plots typical values of the pair
$(y_i,x_i)$ along a DNA segment that contains a homozygous deletion
(copy number~0), a~hemizygous deletion (copy number 1) and a
duplication (copy number 3). Clearly both $y_i$ and $x_i$ convey
information relevant to copy number.

\subsection{Reconstructing a piecewise constant function}\label{fused}

Consider first CNV reconstruction using signal intensities $y_i$ and
neglecting B-allele frequencies $x_i$. While this restriction overlooks
important information, it has the benefit of recasting CNV
reconstruction as a general problem of estimating a piecewise constant
function from linearly ordered observations. In such regression
problems, \citet{FLasso} and \citet{cghFLasso} suggest minimizing the
criterion
\[
f(\bolds{\beta})  =  \frac{1}{2} \sum_{i=1}^n \Biggl(y_i -
\sum_{j=1}^p z_{ij}\beta_j\Biggr)^2 + \lambda_1\sum_{j=1}^p |\beta_j| +
\lambda_2 \sum_{j=2}^p |\beta_j-\beta_{j-1}|.
\]
Here $\mathbf{y}=(y_i)_{n\times1}$ is the response vector,
$\mathbf{Z}=(z_{ij})_{n \times p}$ is the design matrix,
$\bolds{\beta} = (\beta_j)_{n\times1}$ is the parameter vector of
regression coefficients, and $\lambda_1$ and $\lambda_2$ are tuning
parameters that control the sparsity and smoothness of the model. The
model is particularly suited to situations where the number of
regression coefficients $p$ is much larger than the number of cases
$n$. For the special task of CNV detection, we take
$\mathbf{Z}=\mathbf{I}$ (i.e., $p=n$), reducing the objective function
to
\begin{equation}\label{fl}
f(\bolds{\beta})  =  \frac{1}{2} \sum_{i=1}^n (y_i - \beta_i)^2
+ \lambda_1\sum_{i=1}^n |\beta_i| + \lambda_2 \sum_{i=2}^n
|\beta_i-\beta_{i-1}| .
\end{equation}
Notice that $f(\bolds{\beta})$ is strictly convex and coercive, so
a unique minimum exists. When $\lambda_2=0$, the objective function can
be decomposed into a sum of $n$ terms, each depending only on one
$\beta_i$. This makes it very easy to find its minimum using
coordinate descent [\citet{pco};  \citet{wulange}]. Unfortunately, this is not
the case with $\lambda_2 \ne 0$ because the kinks in the objective
function are no longer confined to the coordinate directions.  This
makes coordinate descent much less attractive [\citet{pco}].  Quadratic
programming [\citet{FLasso}; \citet{cghFLasso}] is still available, but its
computational demands do not scale well as $p$ increases.

\begin{figure}

\includegraphics{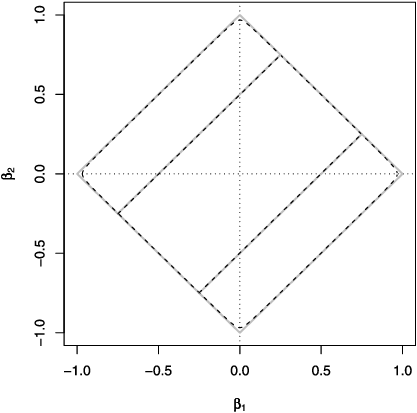}

\caption{Contours corresponding to different penalties. Solid gray
line: $|\beta_1|+|\beta_2|=1$ and $|\beta_1-\beta_2|=\frac{1}{2}$;
Dashed line: $\Vert\beta_1\Vert_{2,\varepsilon}+\Vert\beta_2\Vert_{2,\varepsilon}=1$ and
$\Vert\beta_1-\beta_2\Vert_{2,\varepsilon}=\frac{1}{2}$.}\label{fig2}
\end{figure}

Inspired by the resolution of similar smoothing dilemmas in imaging
[\citet{bioucas06}; \citet{rof}], we simplify the problem by slightly modifying
the penalty. The function
\[
\Vert x \Vert_{2,\varepsilon}  =  \sqrt{x^2+\varepsilon}
\]
is both differentiable and strictly convex.  For small $\varepsilon>0$ it
is an excellent approximation to $|x|$.  Figure \ref{fig2} illustrates
the quality of this approximation for the choice $\varepsilon=0.001$. In
practice, we set $\varepsilon=10^{-10}$.  If we substitute
$\Vert x \Vert_{2,\varepsilon}$ for $|x|$, then the CNV objective function becomes
\begin{equation}\label{substitute_objective}
f_{\varepsilon}(\bolds{\beta})  =  \frac{1}{2} \sum_{i=1}^n (y_i -
\beta_i)^2 + \lambda_1\sum_{i=1}^n \Vert \beta_i\Vert_{2,\varepsilon} + \lambda_2
\sum_{i=2}^n \Vert\beta_i-\beta_{i-1}\Vert_{2,\varepsilon}.
\end{equation}
As $\varepsilon$ tends to 0, one can show that the unique minimum point of
(\ref{substitute_objective}) tends to the unique minimum point of the
original objective function.

Another virtue of the substitute penalties is that they lend themselves
to majorization by a quadratic function. Given the concavity of the
function $t \mapsto \sqrt{t+\varepsilon}$, it is geometrically obvious
that
\[
\Vert x\Vert_{2,\varepsilon}  \leq  \Vert z\Vert_{2,\varepsilon} +
\frac{1}{2\Vert z\Vert_{2,\varepsilon}}[x^2-z^2],
\]
with equality holding if and only if $x=z$. This inequality enables a
Majorization--Minimization (MM) [\citet{LangeOpt}] strategy that
searches for the minimum of the objective function. Each step of this
iterative approach requires the following: (a)~majorizing the objective
function by a surrogate equal to it at the current parameter vector and
(b) minimizing the surrogate. The better-known EM algorithm is a
special case of the MM algorithm. The MM algorithm generates a descent
path guaranteed to lead to the optimal solution when one exists. More
information can be found in \citet{LangeOpt}. Returning to our problem,
we can replace the objective function by the surrogate function
\begin{eqnarray*}
&&g_{\varepsilon,m}\bigl(\bolds{\beta} \mid \bolds{\beta}^{(m)}\bigr) \\
&&\qquad= \frac{1}{2} \sum_{i=1}^n (y_i -\beta_i)^2 + \frac{\lambda_1}{2} \sum_{i=1}^n
\frac{\beta_i^2}{\Vert\beta_i^{(m)}\Vert_{2,\varepsilon}} + \frac{\lambda_2}{2}
\sum_{i=2}^n
\frac{(\beta_i-\beta_{i-1})^2}{\Vert\beta_i^{(m)}-\beta_{i-1}^{(m)}\Vert_{2,\varepsilon}}
+ c_m,
\end{eqnarray*}
where $m$ indicates iteration number and $c_m$ is a constant unrelated
to $\bolds{\beta}$.  Minimization of
$g_{\varepsilon,m}(\bolds{\beta} \mid \bolds{\beta}^{(m)})$ to
obtain $\bolds{\beta} ^{(m+1)}$ drives the objective function
$f_{\varepsilon}(\bolds{\beta})$ downhill. Although the MM algorithm
entails iteration, it replaces the original problem by a sequence of
simple quadratic minimizations.  The descent property of the MM
algorithm guarantees that progress is made every step along the way.
This, coupled with the convexity of our problem, guarantees convergence
to the global minimum.

Despite these gains in simplicity,  the surrogate function still does
not decompose into a sum of $n$ terms, with each depending on only one
$\beta_i$. The fact that the even numbered $\beta_i$ do not interact
given  the odd numbered $\beta_i$  (and vice versa) suggests
alternating updates of the two blocks of even and odd numbered
parameters. In practice, this block relaxation strategy converges too
slowly to be competitive. Fixing $\beta_{i-1}$ and $\beta_{i+1}$ leaves
too little room to move $\beta_{i}$.  Fortunately, full minimization of
the quadratic is less onerous than one might expect.  The surrogate
function can be written in a matrix form
\begin{equation}\label{thomas}
g_{\varepsilon,m}\bigl(\bolds{\beta} \mid \bolds{\beta}^{(m)}\bigr)  =
\tfrac{1}{2} \bolds{\beta}^T \mathbf{A}_m \bolds{\beta} -
\mathbf{b}_m^T\bolds{\beta} + \tilde{c}_m,
\end{equation}
where $\mathbf{A}_m$ is a tridiagonal symmetric matrix.  In view of the
strict convexity of the surrogate function, $\mathbf{A}_m$ is also
positive definite. The nonzero entries of $\mathbf{A}_m$ and
$\mathbf{b}_m$ are \looseness=-1
\begin{eqnarray*}
a_{1,1}^{(m)} & = & 1 +
\frac{\lambda_1}{\Vert\beta_1^{(m)}\Vert_{2,\varepsilon}}
+ \frac{\lambda_2}{\Vert\beta_2^{(m)}-\beta_1^{(m)}\Vert_{2,\varepsilon}}; \\
a_{i,i}^{(m)} & = & 1 +
\frac{\lambda_1}{\Vert\beta_i^{(m)}\Vert_{2,\varepsilon}} +
\frac{\lambda_2}{\Vert\beta_i^{(m)}-\beta_{i-1}^{(m)}\Vert_{2,\varepsilon}}
+ \frac{\lambda_2}{\Vert\beta_{i+1}^{(m)}-\beta_i^{(m)}\Vert_{2,\varepsilon}},\\
&&\hspace{247pt} i=2,\ldots,n-1;\\
a_{n,n}^{(m)} & = & 1 +
\frac{\lambda_1}{\Vert\beta_n^{(m)}\Vert_{2,\varepsilon}}
+ \frac{\lambda_2}{\Vert\beta_n^{(m)}-\beta_{n-1}^{(m)}\Vert_{2,\varepsilon}}; \\
a_{i,i+1}^{(m)} & = & -\frac{\lambda_2}{\Vert\beta_{i+1}^{(m)}-\beta_i^{(m)}\Vert_{2,\varepsilon}}, \qquad i=1,\ldots,n-1; \\
a_{i-1,i}^{(m)} & = & -\frac{\lambda_2}{\Vert\beta_{i}^{(m)}-\beta_{i-1}^{(m)}\Vert_{2,\varepsilon}},\qquad  i=2,\ldots,n; \\
b_i^{(m)} & =&   y_i, \qquad i=1,\ldots,n.
\end{eqnarray*}
The minimum of the quadratic occurs at the point $\bolds{\beta} =
\mathbf{A}_m^{-1}\mathbf{b}_m$.  Thanks to the simple form of
$\mathbf{A}_m$, there is a variant of Gaussian elimination known as the
tridiagonal matrix algorithm (TDM) or Thomas's algorithm [\citet{TDMA}]
that solves the linear system
$\mathbf{A}_m\bolds{\beta}=\mathbf{b}_m$ in just $9n$ floating
point operations. Alternatively, one can exploit the fact that the
Cholesky decomposition of a banded matrix is banded with the same
number of bands. As illustrated in Section \ref{MMTDM_vs_MMB}, Thomas's
algorithm is a vast improvement over block relaxation.

A few comments on the outlined strategy are in order. By changing the
penalty from $\|\cdot\|_{\ell_1} $ to $\Vert\cdot\Vert_{2,\varepsilon} $, we
favor less sparse solutions. However,  spareness is somewhat besides
the point. What we really need are criteria for calling deletions and
duplications.  The lasso penalty is imposed in this problem because
most chromosome regions have a normal copy number where $y_i$ hovers
around 0. The same practical outcome can be achieved by imputing copy
number 2 for regions where the estimated $\beta_i$ value is close to 0.
(See Section \ref{fdr}.) It is also relevant to compare our
minimization strategy to that of \citet{pco}. The fusion step of their
algorithm has the advantage of linking coefficients that appear to be
similar, but it has the disadvantage that once such links are forged,
they cannot be removed. This permanent commitment may preclude finding
the global minimum,
 a limitation that our MM algorithm does not share.

Perhaps more importantly, our strategy can be adapted to handle more
general objective functions, as long as the resulting matrix
$\mathbf{A}$ in (\ref{thomas}) is banded, or, at least, sparse. For
example, consider the inpainting problem in image reconstruction
[\citet{inpainting}]. In this two dimensional problem, the intensity
levels for certain pixels are lost. Let $S$ be the set of pixels with
known levels.  The objective function
\begin{eqnarray*}
f(\bolds\beta) & = & \frac{1}{2}\sum_{(i,j) \in S} (y_{ij}-\beta_{ij})^2 \\
&& {}+ \lambda \sum_{i=1}^n\sum_{j=2}^n
\Vert\beta_{ij}-\beta_{i,j-1}\Vert_{2,\varepsilon} + \lambda
\sum_{i=2}^n\sum_{j=1}^n \Vert\beta_{ij}-\beta_{i-1,j}\Vert_{2,\varepsilon}
\end{eqnarray*}
represents a compromise between imputing unknown values and smoothing.
If we majorize the penalties in this objective function by quadratics,
then we generate a quadratic surrogate function.  The corresponding
Hessian of the surrogate is very sparse. (Actually, it is banded, but
not in a useful fashion.) Although we can no longer invoke Thomas's
algorithm, we can solve the requisite system of linear equations by a
sparse conjugate gradient algorithm.

All of the algorithms mentioned so far rely on known values for the
tuning constants. We will describe our operational choices for these
constants after discussing the problem of imputing chromosome states
from estimated parameters in the next section.

\subsection{Reconstructing discrete copy number states}

Imputation of copy number as just described has the drawbacks of
neglecting relevant information and requiring the estimation of a large
number of parameters. To overcome these limitations, we now bring in
the BAF $x_i$ and focus on a model with a finite number of states. This
setting brings us much closer to the HMM framework, often used for CNV
reconstruction. Such similarity will be evident also in the numerical
strategy we will use for optimization. However, our approach avoids the
distributional assumptions  at the basis of an HMM.

We consider 10 possible genotypic states $\phi$, A, B, AA, AB, BB, AAA,
AAB, ABB and BBB at each SNP.  Here $\phi$ is the null state with a
copy number of 0.  (Note that, in the interest of parsimony,  we
contemplate double deletions, but not double duplications. This has
more to do with the strength of signal from duplications than their
actual frequency, and it is an assumption that can be easily relaxed.)
In the model the average signal intensity $\mu_{c(s)}$ for a state $s$
depends only on its copy number $c(s)$. Regardless of whether we
estimate the $\mu_c$ or fix them, they provide a more parsimonious
description of the data than the  $\beta_i$, which could take on a
different value for each SNP. Furthermore, while we still need to
impute a  state for each SNP $i$, selecting one possible value out of
10 is intrinsically easier than estimation of the continuously varying
$\beta_i$. Table \ref{table1} lists the copy number $c(s)$, the
expected value of $y_i$ and the approximate distribution of $x_i$ for
each genotype state~$s$. To reconstruct the state vector
$\mathbf{s}=(s_1,\ldots, s_n)$, we recommend minimizing the generic
objective function
\begin{eqnarray}\label{dina}
f(\mathbf{s}) & = &  \sum_{i=1}^n L_1(y_i,s_i)+\alpha\sum_{i=1}^n L_2(x_i,s_i)\nonumber
\\ [-8pt]\\ [-8pt]
& &{} +\lambda_1\sum_{i=1}^n\big|\mu_{c(s_i)}\big| +\lambda_2
\sum_{i=2}^{n}\big|\mu_{c(s_{i})}-\mu_{c(s_{i-1})}\big|, \nonumber
\end{eqnarray}
which again is a linear combination of losses plus penalties. Here
$\alpha$, $\lambda_1$ and $\lambda_2$ are positive tuning constants
controlling the relative influences of the various factors. The lasso
penalty makes the states with copy number 2 privileged. The fused-lasso
penalty discourages changes in state. Minimizing the objective function
(\ref{dina}) is a discrete rather than a continuous optimization
problem.

\begin{table}\tablewidth=285pt
\tabcolsep=0pt
\caption{Genotype states, corresponding copy numbers, expected values
of $y_i$ and approximate distributions of $x_i$} \label{table1}
\begin{tabular*}{285pt}{@{\extracolsep{\fill}}lcf{3.6}k{2.3}@{}}
\hline
\textbf{Genotype state} $\bolds{s}$ & \textbf{Copy number} $\bolds{c(s)}$ & \multicolumn{1}{c}{\textbf{Mean of} $\bolds{y_i}$}  & \multicolumn{1}{c}{\textbf{Distribution of} $\bolds{x_i}$} \\
\hline
$\phi$ & 0 & \mu_0.(<\mu_1) & \multicolumn{1}{c}{Uniform on $[0,1]$}\\
  A  & 1 & \mu_1.(<0) & \approx.0\\
  B  & 1 & \mu_1.(<0) & \approx.1\\
  AA & 2 & \mu_2.(\approx 0) & \approx.0\\
  AB & 2 & \mu_2.(\approx 0) & \approx.1/2\\
  BB & 2 & \mu_2.(\approx 0) & \approx.1\\
 AAA & 3 & \mu_3.(>0) & \approx.0\\
 AAB & 3 & \mu_3.(>0) & \approx.1/3\\
 ABB & 3 & \mu_3.(>0) & \approx.2/3\\
 BBB & 3 & \mu_3.(>0) & \approx.1\\
\hline
\end{tabular*}
\end{table}

Different loss functions may be appropriate in different circumstances.
If the intensity values are approximately Gaussian around their means
with a common variance, then the choice $L_1(y,s)=[y-\mu_{c(s)}]^2$ is
reasonable. For the BAF $x_i$, the choice $L_2(x,s)=(x-\nu_s)^2$ is
also plausible. Here $\nu_s$ is the centering constant appearing in the
fourth column of Table \ref{table1}. For instance,
$L_2(x,\mbox{ABB})=(x-2/3)^2$. For the null state $\phi$, we would take
\[
L_2(x,\phi)  =  \int_0^1 (x-u)^2\,du =
\frac{1}{3}[x^3+(1-x)^3].
\]

Once the loss functions are set, one can employ dynamic programming to
find the state vector $\mathbf{s}$ minimizing the objective function
(\ref{dina}).  If we define the partial solutions
\[
g_i(j)  =  \min_{s_1,\ldots,s_{i-1}} f(s_1,\ldots,s_{i-1},s_i=j)
\]
for $i=1,\ldots,n$, then the optimal value of the objective function is
$\min_{j} g_{n}(j)$. We evaluate the partial solutions $g_i(j)$
recursively via the update
\begin{eqnarray}\label{rec}
g_{i+1}(j) & = & \min_k \bigl[g_i(k) + L_1(y_{i+1},j)+\alpha L_2(x_{i+1},j) \\
&&\hspace{37pt} {}+ \lambda_1 \big|\mu_{c(j)}\big| +\lambda_2
\big|\mu_{c(j)}-\mu_{c(k)}\big|\bigr], \nonumber
\end{eqnarray}
with initial conditions
\[
g_1(j)  =  L_1(y_1,j)+\alpha L_2(x_1,j) +\lambda_1 \big|\mu_{c(j)}\big|.
\]
The beauty of dynamic programming is that it applies to a variety of
loss and penalty functions.

In fact, it is possible to construct an even more parsimonious model
whose four states correspond to the four copy numbers 0, 1, 2 and 3.
The loss function $L_1(y,c)=(y-\mu_c)^2$ is still reasonable, but
$L_2(x,c)$ should reflect the collapsing of genotypes. Here $c$ is the
copy number. Two formulations are particularly persuasive. The first
focuses on the minimal loss among the genotypes relevant to each copy
number. This produces
\begin{equation}\label{mle}
\quad L_2(x,c) = \cases{
\displaystyle\int_0^1 (x-u)^2\,du = \frac{1}{3}[x^3+(1-x)^3], & \quad $c=0$,
\cr
\min \{(x-0)^2, (x-1)^2\}, & \quad $c=1$, \cr
\min \{(x-0)^2,(x-1/2)^2,(x-1)^2\}, & \quad $c=2$, \cr
\min \{(x-0)^2,(x-1/3)^2,(x-2/3)^2,(x-1)^2\}, &\quad $c=3$.
}
\end{equation}

The second formulation averages loss weighted by genotype frequency.
There are other reasonable loss functions.  Among these it is worth
mentioning negative log-likelihood, Huber's function and the hinge loss
function of machine learning.

Dynamic programming does require specification of the parameters
characterizing the distribution of the intensities $y_i$ and the BAF
$x_i$. It may be possible to assign values to these parameters based on
previous data analysis. If not, we suggest estimating them concurrently
with assigning states. For example, if the parameters are the intensity
means $\mu_0$, $\mu_1$, $\mu_2$ and $\mu_3$, then, in practice, we
alternate two steps starting from plausible initial values for the
$\mu_i$. The first step reconstructs the state vector $\mathbf{s}$.
The second step re-estimates the $\mu_i$ conditional on these
assignments.  Thus, if $G_i$ is the group of SNPs assigned copy number
$i$, then we estimate $\mu_i$ by the mean of the $y_i$ over $G_i$.
Taking the median rather the mean makes the process robust to outliers.
A few iterations of these two steps usually gives stable parameter
estimates and state assignments. To further stabilize the process, we
impose two constraints on the second step. If the number of SNPs
assigned to $G_i$ is less than a threshold, say, $5$, we choose not to
update $\mu_i$ and rather keep the estimate in the previous iteration.
In each update we enforce the order of $\mu_0 < \mu_1 < \mu_2 (\approx
0) < \mu_3$. In the following we will refer to the approach described
in this section as dynamic programming imputation (DPI).

\section{Results}

\subsection{Identification of deleted and duplicated segments by the fused lasso} \label{fdr}

In calling deletions and duplications with the fused lasso, we adopt
the procedure of \citet{cghFLasso}. Originally designed for array-CGH
platforms, this procedure aims to control false discovery rate (FDR).
Fortunately, it can be readily applied to genotype data. The general
idea is to formulate the problem as one of multiple hypothesis testing
for nonoverlapping chromosome segments $S_1$ through~$S_K$.  For each
segment $S_k$ we define the test statistic
\[
\hat{z}_k  =  \frac{\sum_{i \in
S_k}\hat{\beta}_i}{\sqrt{n_k}\hat{\sigma}},
\]
where $n_k$ is the number of SNPs in segment $S_k$ and $\hat{\sigma}$
is a conservative estimate of standard deviation of the $\hat{\beta}_i$
across all segments based on the $y_i$ values between their 2.5 and
97.5 percentiles. The associated $p$-value for segment $S_k$ is
approximated by $p_k=2P(Z>|\hat{z}_k|)$ for $Z \sim \mathcal{N}(0,1)$.
For a given threshold $q \in (0,1)$, we estimate the FDR by
\begin{equation}\label{estimate_FDR}
\widehat{\mbox{FDR}}(q)   =   \frac{Kq \cdot
1/K\sum_{k=1}^K n_k} {\sum_{k=1}^K n_k 1_{(p_k \leq q)}}  =
 \frac{q\sum_{k=1}^K n_k} {\sum_{k=1}^K n_k 1_{(p_k \leq q)}}.
\end{equation}
Here the FDR is defined as the ratio between the number of SNPs in
nominal CNV segments with true copy number $2$ and the total number of
SNPs claimed to be within CNV segments. In the FDR estimate
(\ref{estimate_FDR}), $q$ is roughly regarded as the fraction of null
(copy number $2$) segments among all candidate CNV segments. In the
numerator, $\frac{1}{K}\sum_{k=1}^K n_k$ counts the average SNP number
within each segment, and $Kq$ estimates the expected number of null
segments.  In the denominator, $\sum_{k=1}^K n_k 1_{(p_k \leq q)}$
counts the total number of SNPs claimed to be located in CNV segments.
Thus, this approximation is desired according to the SNP-number-based
definition.

Once we decide on an FDR level $\alpha$, the threshold $q$ is
determined as the largest value satisfying $\widehat{\mbox{FDR}}(q)
\leq \alpha$. We call a segment $S_k$ a deletion if $\hat{z}_k < 0$ and
$p_k \leq q$ and a duplication if $\hat{z}_k > 0$ and $p_k \leq q$.

\subsection{Choice of tuning constants}

Choice of the tuning constants $\lambda_1$ and $\lambda_2$ is
nontrivial.  Because they control the sparsity and smoothness of the
parameter vector $\bolds{\beta}$ and therefore drive the process
of imputation, it is crucial to make good choices. Both of the
references \citet{pco} and \citet{wulange} discuss the problem and
suggest solutions in settings similar to ours. While explicit
theoretical expressions for optimal $\lambda_1$ and $\lambda_2$ are
currently unavailable, known results can inform practical choices.

\citet{pco} consider the optimal solution to the fused-lasso problem
\[
\hat{\bolds{\beta}}(\lambda_1,\lambda_2) =
\arg\min_{\bolds{\beta}} \frac{1}{2} \sum_{i=1}^n (y_i -
\beta_i)^2 + \lambda_1\sum_{i=1}^n |\beta_i| + \lambda_2 \sum_{i=2}^n
|\beta_i-\beta_{i-1}|.
\]
They prove that $\hat{\bolds{\beta}}(\lambda_1,\lambda_2)$ for
$\lambda_1>0$ is a soft-thresholding of
$\hat{\bolds{\beta}}(0,\lambda_2)$ when\hspace*{-0.45pt}\break$\lambda_1=0$, namely,
\begin{equation}\label{tl}
\hat{\beta}_i(\lambda_1,\lambda_2) =
\sign(\hat{\beta}_i(0,\lambda_2))\bigl(|\hat{\beta}_i(0,\lambda_2)|-\lambda_1\bigr)_{+},
\qquad i = 1,\ldots,n.
\end{equation}
This implies that $\lambda_1>0$ will drive to 0 those segments of the
piecewise constant solution $\hat{\bolds{\beta}}(0,\lambda_2)$
whose absolute values are close to 0. It is also important to note
that, since $\hat{\bolds{\beta}}(0,\lambda_2)$ is piecewise
constant, its effective dimension is much lower than $n$.

To understand how the optimal values of these tuning parameters depend
on the dimension of the vector $\bolds{\beta}$, let us recall
pertinent properties of the Lasso estimator in linear regression. In
this setting
\begin{equation} \label{reg}
\hat{\bolds{\beta}} =\arg\min_{\bolds{\beta}}
\frac{1}{2} \Vert\mathbf{y}-\mathbf{Z}\bolds{\beta}\Vert_{\ell_2}^2 +
\lambda\Vert\bolds{\beta}\Vert_{\ell_1},
\end{equation}
where $\mathbf{y}_{n \times 1} \sim {\mathcal N}(\mathbf{Z}_{n \times
p}\bolds{\beta}_{p \times 1},\sigma^2\mathbf{I}_{n \times n})$,
and $\Vert\cdot\Vert_{\ell_1}$ and $\Vert\cdot\Vert_{\ell_2}$ are the $\ell_1$ and
$\ell_2$ norms. \citet{Candes}, \citet{Donoho} and \citet{Wainwright} show
that a Lasso estimator with $\lambda=c\sigma\sqrt{\log p}$ for some
constant $c$ leads to an optimal upper bound on
$\Vert\mathbf{Z}\bolds{\beta} -
\mathbf{Z}\hat{\bold{\beta}}\Vert_{\ell_2}^2 $. Our problem with $
\lambda_1=0$ fits in this framework if we reparameterize via $\delta_1
= \beta_1$ and $\delta_i = \beta_i-\beta_{i-1}$ for $i=2,\ldots,n$. In
the revised problem
\begin{equation}\label{repar}
\hat{\bolds{\delta}} = \arg\min_{\bolds{\delta}}
\frac{1}{2} \sum_{i=1}^n \Biggl(y_i - \sum_{j=1}^i\delta_j\Biggr)^2 +
\lambda_2\sum_{i=2}^n |\delta_i|,
\end{equation}
$p=n$, and the design matrix is lower-triangular with all nonzero
entries equal to~$1$. This finding suggests that we scale $\lambda_2$
by $\sqrt{\log n}$.

On the basis of these observations, we explored the choices
\[
\lambda_1 = \rho_1 \sigma,\qquad \lambda_2 = \rho_2
\sigma\sqrt{\log n},\qquad\rho_1,\rho_2>0.
\]
Here $\sigma$ relates the tuning parameters to the noise level. Because
the effective dimension in (\ref{tl}) is much smaller than $n$, we
assumed that $\lambda_1$ does not depend on $n$. Although $\rho_1$ and
$\rho_2$ can be tuned more aggressively by cross-validation or criteria
such as BIC, we chose the sensible and operational combination
\begin{equation}\label{L1L2choice}
\lambda_1 =  \sigma,\qquad \lambda_2  =  2\sigma\sqrt{\log n}.
\end{equation}
A small scale simulation study suggested that the performance of our
methods does not vary substantially for values of $\rho_1$ and $\rho_2$
close to 1 and 2, respectively. One may also vary $\rho_1$ and/or
$\rho_2$ mildly to achieve different combinations of sensitivity and
specificity as defined in Section \ref{accuracy_index}. (Data not
shown.)

In practice, we do not know the value of $\sigma$. Here we estimated a
different $\sigma$ for each individual, using the standard deviation of
$y_i$ values between their $2.5$ and $97.5$ percentiles. We decided to
use only data points within the $95\%$-interquantile range in order to
exclude values of $y_i$ corresponding to possible deletions and
duplications. Other possible robust estimators are based on the median
absolute deviation or the winsorized standard deviation. In a
small-scale simulation we did not observe substantial differences
between these estimators. (Data not shown.)

While most of the experiments in the paper used the values of
$\lambda_1$ and $\lambda_2$ suggested in equation (\ref{L1L2choice}),
we also designed and conducted a more general simulation study to find
the optimal values of these tuning parameters; see Section
\ref{accuracy_by_length} for details.

\subsection{Simulated data with {\it in silico} CNVs}

To illustrate the effectiveness of our algorithms, we tested them on
simulated data. Real data with empirically validated CNVs would be
ideal, but such a gold standard does not exist. Instead, we used data
from male and female X chromosomes to construct
 \textit{in silico} CNV.  Since males are equipped with only one X chromosome, we can use their genotype data to approximate the signal generated by deletion regions.  A patchwork of female and male data mimics what we expect from an ordinary pair of homologous chromosomes with occasional deletions. Our X chromosome data come from the schizophrenia study sample of \citet{schizoAJHG08} genotyped on the Illumina platform.  We focus on the $307$ male and $344$ female controls.

To avoid artifacts, the data needed to be preprocessed. We identified
SNP clusters on the X chromosome using the Beadstudio Illumina software
on female controls. These clusters permit estimation of parameters
typical of a diploid genome. We then normalized the corresponding male
SNP signals relative to the corresponding female signals.  Finally, to
destroy the signature of possible  CNVs in the female data, we permuted
the order of the SNPs. This action breaks up the patterns expected
within CNV regions and eliminates the smooth variation in the intensity
signals [\citet{gcadj}].

After these preprocessing steps, we generated ordinary copy number
regions from the female data and deleted regions from the male data.
We also generated duplications by taking the weighted averages
\begin{eqnarray*}
y_{i,dup} & = & y_{i,f} + 0.55 \times |\mbox{median}(y_f) - \mbox{median}(y_m)|, \\
x_{i,dup} & = & \tfrac{1}{3}x_{i,m} + \tfrac{2}{3}x_{i,f}
\end{eqnarray*}
for the intensities and BAFs, where the $f$ and $m$ subscripts refer to
females and males. Because duplications show a lesser increase in logR
values than the deletions show a decrease, the factor 0.55 multiplies
the absolute difference $|\mbox{median}(y_f)-\mbox{median}(y_m)|$
between median female and male intensities.

We generated two different data sets to assess the operating
characteristics of the proposed algorithms. In both data sets the
number of deletions equals the number of duplications. \textit{Data set 1}
consists of 3600 sequences, each 13,000 SNPs long, with either a
deletion or a duplication in the central position.  The CNVs had
lengths evenly distributed over the 6 values 5, 10, 20, 30, 40 and 50
SNPs.  \textit{Data set 2 } consists of 300 sequences with variable
numbers of SNPs and either a deletion or duplication in the central
position.  The sequence lengths were evenly distributed over the values
4000, 8000, 12,000, 16,000 and 20,000 SNPs; the CNV lengths followed the
distribution of data set 1.

The sequence and CNV lengths in our simulations were chosen to roughly
mimic values expected in real data. For the Illumina HumanHap550
BeadChip platform, the median number of SNPs per chromosome arm is
13,279, with a median absolute deviation of $8172$. Current empirical
data suggests that there is usually at most one CNV per chromosome arm
[\citet{pennCNV}] and that the length of the typical CNV is usually less
than $50$ SNPs [\citet{HapMapCNV}]. The sequences from data set 1
represent an average chromosome arm, while the sequences from data set
2 capture the diversity across all chromosome arms. Both data sets have
useful lessons to teach.

\subsection{Measures of accuracy and a benchmark algorithm}  \label{accuracy_index}

We will measure accuracy on a SNP by SNP basis,  adopting the following
indexes: true positive rate (TPR or sensitivity), false positive rate
(FPR or 1-specificity), and false\vadjust{\goodbreak} discovery rate (FDR). These are
defined as the ratios
\begin{eqnarray*}
\mbox{TPR} & = & \frac{\mbox{TP}}{\mbox{P}} =  \frac{\mbox{TP}}{\mbox{TP}+\mbox{FN}}, \\
\mbox{FPR} & = & \frac{\mbox{FP}}{\mbox{N}}  =  \frac{\mbox{FP}}{\mbox{FP}+\mbox{TN}}, \\
\mbox{FDR} & = & \frac{\mbox{FP}}{\mbox{TP}+\mbox{FP}},
\end{eqnarray*}
where the capital letters T, F, P, N and R stand for true, false,
positive, negative and rate, respectively. For example, the letter P by
itself should be interpreted as the number of SNPs with true copy
number equal to 0, 1 or 3; the pair of letters FN should be interpreted
as the number of SNPs with true copy number 0, 1 or 3 but imputed copy
number 2. We will also evaluate the number of iterations until
convergence and the overall computational time required by each
algorithm.

For benchmarking purposes, we will compare the performance of the
proposed algorithms to that of PennCNV [\citet{pennCNV}], a
state-of-the-art hidden Markov model for CNV discovery on Illumina
data. PennCNV bases the genotype call for SNP $i$ on its $y_i$ and
$x_i$ measurements and its major and minor allele frequencies. We
expect PennCNV to perform well because it has been extensively tuned on
real and simulated data. The main aim of our comparisons is simply to
check whether the new algorithms suffer a substantial loss of accuracy
relative to PennCNV.

\subsection{Convergence of the MMTDM and MMB algorithms} \label{testTDM}  \label{MMTDM_vs_MMB}

\begin{figure}[b]
\tabcolsep=0pt
\begin{tabular*}{\textwidth}{@{\extracolsep{\fill}}cc@{}}

\includegraphics{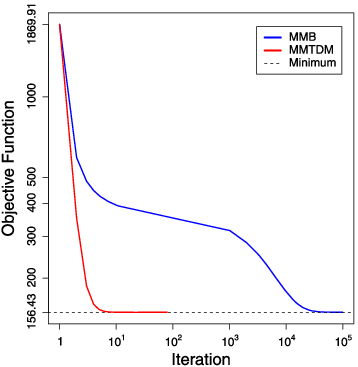}
&\includegraphics{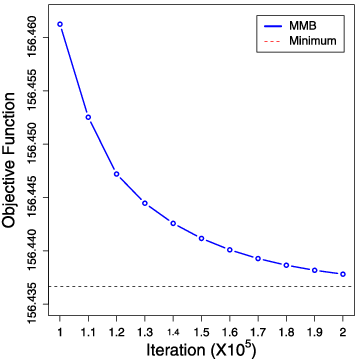}\\
(a)& (b)\\ [7 pt]
\end{tabular*}
\caption{Comparison of convergence rates for the two algorithms MMB
and MMTDM for the fused lasso. \textup{(a)} MMTDM converges much faster than
MMB. Blue line: MMB; Red line: MMTDM; Black dashed line: minimum value
of objective function; \textup{(b)} After $10^5$ iterations, MMB converges with
an accuracy of $0.01$.}\label{fig3}
\end{figure}

We first investigate two versions of the fused-lasso procedure. Both
implement the MM algorithm on the objective function
(\ref{substitute_objective}). The MMTDM algorithm solves the
minimization step by the tridiagonal matrix algorithm.  The MMB
algorithm approximately solves the minimization step by one round of
block relaxation. To assess the rate of convergence of MMTDM and MMB,
we used data set 1 with 3600 sequences of 13,000 SNPs each.  We declared
convergence for a run when the difference between the objective
function at two consecutive iterations fell below $10^{-4}$. To limit
the computational burden, we set the maximum number of iterations equal
to 10,000. Both algorithms started with the values $\beta_i = y_i$.
Each entry of Table \ref{table2} summarizes the results for a different
CNV width. The table makes it abundantly clear that MMB is not
competitive. Because MMB never converged in these trials, we took one
sequence and ran it to convergence under the more stringent convergence
criterion of $10^{-6}$.  Figure \ref{fig3} plots the value of the
objective function under the two algorithms. Examination of these plots
shows that MMTDM is on the order of 1000 times faster than MMB.

\begin{table}
\tabcolsep=0pt
\caption{Number of iterations until convergence of MMTDM and MMB. For
MMTDM, each entry summarizes the average number of iterations required
for convergence; Standard errors appear in parentheses. MMB never
converges within 10,000 iterations in this case} \label{table2}
\begin{tabular*}{\textwidth}{@{\extracolsep{\fill}}lcccccc@{}}
\hline
\textbf{CNV size} & \hspace*{9pt}\textbf{5} & \hspace*{8pt}\textbf{10} & \hspace*{9pt}\textbf{20} & \hspace*{9pt}\textbf{30}
& \hspace*{9pt}\textbf{40} &\hspace*{7pt} \textbf{50} \\
\hline
MMB & $>$10,000 & $>$10,000 & $>$10,000
& $>$10,000 & $>$10,000 & $>$10,000 \\
MMTDM & \phantom{000000}33.1\hspace*{-3pt} & \phantom{000000}33.3\hspace*{-3pt} & \phantom{000000}34.5\hspace*{-3pt} & \phantom{000000}33.3\hspace*{-3pt} & \phantom{000000}33.7\hspace*{-3pt} & \phantom{000000}33.9\hspace*{-3pt}  \\
 & \phantom{000000}(13.0)\hspace*{-3pt} & \phantom{000000}(12.0)\hspace*{-3pt} & \phantom{000000}(13.9)\hspace*{-3pt} & \phantom{000000}(12.9)\hspace*{-3pt} & \phantom{000000}(12.2)\hspace*{-3pt} & \phantom{000000}(12.1)\hspace*{-0.3pt} \\
\hline
\end{tabular*}
\end{table}

\subsection{Effect of including BAF in discrete reconstruction}

Data set 1 also illustrates the advantages of including BAF information
in CNV reconstruction. Here we focus on dynamic programming imputation
(DPI) based on the objective function (\ref{dina}). Note that this
function does not incorporate prior knowledge of the frequency of
deletions versus duplications. In running the dynamic programming
algorithm, we rely on  results from a previous study [\citet{us}] to
initialize the intensity parameters $\mu_k$. Because the $\mu_k$ are
re-estimated after each round of imputation, we can safely ignore the
slight differences between the genotyping platforms of the previous and
current studies. Table \ref{table3} reports the various accuracy
indexes as a function of the tuning constant $\alpha$ determining the
relative influence of BAF. Although we already have acceptable
reconstruction for $\alpha=0$, increasing it leads to substantial
improvements. When $\alpha=12$, we reach an excellent balance between
sensitivity and specificity. In the following we adopt the value
$\alpha=12$ unless noted to the contrary.

\begin{table}
\tabcolsep=0pt
\caption{TPR, FPR and FDR in DPI as $\alpha$ varies} \label{table3}
\begin{tabular*}{\textwidth}{@{\extracolsep{\fill}}lcccllccc@{}}
\hline
$\bolds\alpha$ & \textbf{TPR (\%)} & \textbf{FPR (\%)} & \textbf{FDR (\%)} && $\bolds\alpha$ & \textbf{TPR (\%)} & \textbf{FPR (\%)} & \textbf{FDR (\%)} \\
\hline
  \phantom{0}0 & 87.56 & 0.0064 & 3.53 && 15 & 94.08 & 0.0010 & 0.53 \\
   \phantom{0}1 & 89.55 & 0.0031 & 1.70 && 16 & 94.14 & 0.0010 & 0.53 \\
   \phantom{0}2 & 90.68 & 0.0019 & 1.04 &&17 & 94.18 & 0.0010 & 0.54 \\
   \phantom{0}3 & 91.57 & 0.0017 & 0.92 && 18 & 94.22 & 0.0011 & 0.57 \\
   \phantom{0}4 & 92.14 & 0.0014 & 0.77 && 19 & 94.26 & 0.0011 & 0.57 \\
   \phantom{0}5 & 92.55 & 0.0012 & 0.63 && 20 & 94.30 & 0.0012 & 0.63 \\
   \phantom{0}6 & 92.80 & 0.0010 & 0.53 && 21 & 94.37 & 0.0012 & 0.65 \\
  \phantom{0}7 & 93.06 & 0.0010 & 0.53 && 22 & 94.39 & 0.0013 & 0.68 \\
   \phantom{0}8 & 93.27 & 0.0010 & 0.52 && 23 & 94.46 & 0.0015 & 0.77 \\
   \phantom{0}9 & 93.50 & 0.0010 & 0.51 && 24 & 94.48 & 0.0015 & 0.81 \\
10 & 93.58 & 0.0009 & 0.49 && 25 & 94.50 & 0.0016 & 0.83 \\
11 & 93.66 & 0.0009 & 0.50 && 26 & 94.53 & 0.0016 & 0.86 \\
12 & 93.83 & 0.0009 & 0.49 && 27 & 94.55 & 0.0018 & 0.93 \\
13 & 93.94 & 0.0009 & 0.49 && 28 & 94.62 & 0.0018 & 0.95 \\
14 & 94.02 & 0.0010 & 0.52 && 29 & 94.59 & 0.0019 & 1.02 \\
\hline
\end{tabular*}
\end{table}

\subsection{Accuracy comparisons for various CNV sizes}
Table \ref{table4} reports the values of the accuracy indices for
various CNV sizes and types. Here we compare PennCNV, fused-lasso
minimization under MMTDM and DPI on data set 1. To avoid overfitting
and a false sense of accuracy, we used $3$-fold cross-validation to
choose $\alpha$. The accuracy indices reported in the table represent
averages over the left-out thirds. Although PennCNV falters a little
with the shortest CNVs, it is clearly the best of the three methods.
More surprising, DPI achieves comparable FPR and FDR to PennCNV as well
as fairly good TPR. In particular, its FDR is uniformly low across CNV
sizes and types. Overall, Table \ref{table4} demonstrates the promise
of DPI. In contrast, the results for fused-lasso minimization are
discouraging.  Despite its post-processing to control FDR, it does
poorly in this regard.  Furthermore, it displays substantially worse
TPR for duplications than PennCNV and DPI, particularly for
duplications spanning only 5 SNPs. This behavior is to be expected
given the poor ability of signal strength alone to separate
duplications from normal chromosome regions. The performance of
fused-lasso minimization underscores the advantages of explicitly
modeling the discrete nature of the state space and taking BAF
information into account. Nonetheless, it is important to keep in mind
that the previous data sets are by design more favorable to PennCNV and
DPI. The analysis of tumor samples with ambiguous copy numbers or
signals from experimental devices, such as CGH arrays that lack
allele-specific information, are bound to cast fused-lasso minimization
in a kinder light.

\begin{table}
\caption{Accuracy comparison of three methods for various CNV sizes.
All accuracy indexes are listed as percentages. The average tuning
parameters used in the fused lasso were $\lambda_1=0.13\,(0.04)$ and
$\lambda_2=0.77\,(0.22)$; standard deviations appear in parentheses. For
DPI, the 3-fold cross-validation accuracy indexes are averages over the
leftover thirds; initial values of average LogR for each copy number
state: $\mu_0 = -5.5923$, $\mu_1 = -0.6313$, $\mu_2 = -0.0045$, $\mu_3
= 0.3252$} \label{table4}
\begin{tabular*}{\textwidth}{@{\extracolsep{\fill}}llccccccccc@{}}
\hline
 && \multicolumn{3}{c}{\textbf{PennCNV}} & \multicolumn{3}{c}{\textbf{Fused Lasso}} &
\multicolumn{3}{c@{}}{\textbf{DPI}}\\ [-7pt]
&&\multicolumn{3}{c}{\hrulefill}&\multicolumn{3}{c}{\hrulefill}&\multicolumn{3}{c@{}}{\hrulefill}\\
\multirow{2}{19pt}[6pt]{\textbf{CNV size}} &  \multirow{2}{19pt}[6pt]{\textbf{CNV type}} &\textbf{TPR} & \textbf{FPR}  & \textbf{FDR} &  \textbf{TPR} & \textbf{FPR}  & \textbf{FDR} & \textbf{TPR} & \textbf{FPR}  & \textbf{FDR}
\\
\hline
 \phantom{0}5 & Del  & 83.80 & 0.0017 & 4.92 & 76.67 & 0.0202 & 40.66 & 76.67 & 0.0006 & 1.88 \\
     & Dup & 58.53 & 0.0011 & 4.67 & 00.33 & 0.0065 & 98.05 & 53.60 & 0.0003 & 1.28 \\
10 & Del  & 95.03 & 0.0011 & 1.45 & 94.23 & 0.0130 & 15.21 & 89.37 & 0.0005 & 0.77 \\
     & Dup & 93.43 & 0.0006 & 0.78 & 26.00 & 0.0128 & 39.01 & 92.30 & 0.0006 & 0.89 \\
20 & Del  & 94.63 & 0.0008 & 0.58 & 96.97 & 0.0159 & 09.62 & 89.87 & 0.0016 & 1.15 \\
     & Dup & 96.13 & 0.0014 & 0.92 & 74.93 & 0.0126 & 09.86 & 95.50 & 0.0011 & 0.76 \\
30 & Del  & 94.57 & 0.0006 & 0.28 & 96.76 & 0.0156 & 06.53 & 94.73 & 0.0013 & 0.62 \\
     & Dup & 96.09 & 0.0001 & 0.05 & 85.84 & 0.0173 & 08.02 & 95.39 & 0.0012 & 0.55 \\
40 & Del  & 97.83 & 0.0018 & 0.59 & 98.33 & 0.0158 & 04.94 & 98.46 & 0.0006 & 0.19 \\
     & Dup & 94.61 & 0.0014 & 0.46 & 87.88 & 0.0181 & 06.24 & 94.66 & 0.0012 & 0.42 \\
50 & Del & 94.33 & 0.0003 & 0.07 & 95.49 & 0.0162 & 04.21 & 93.82 & 0.0010 & 0.26 \\
     & Dup & 94.50 & 0.0003 & 0.09 & 91.06 & 0.0121 & 03.33 & 95.03 & 0.0011 & 0.30
     \\ [5pt]
\multicolumn{2}{@{}l}{Overall} & 94.42 & 0.0009 & 0.49 & 88.00 & 0.0147 & 07.73 & 93.70 & 0.0009 & 0.50 \\
\hline
\end{tabular*}
\end{table}

\subsection{Accuracy comparison for various SNP sequence lengths} \label{accuracy_by_length}

Data set 2 allowed us to assess performance on longer sequences with
less frequent SNPs and to gain insight into the impact of the tuning
parameters $\lambda_1$ and $\lambda_2$. For the latter purpose we
adopted two strategies: (a) define $\lambda_1$ and $\lambda_2$ by the
values displayed in equation (\ref{L1L2choice}), and (b) adopt an
``oracle'' approach that relies on the knowledge of locations of
deletions and duplications.  Strategy (b) chooses constant  values
across the individuals to maximize TPR (sensitivity) while keeping FPR
and FDR levels comparable to those under strategy (a). The oracle
approach is not applicable to real data sets, where locations of
deletions and duplications are unknown. We adopted it in this analysis
to determine how optimal tuning parameters vary with sequence length.

Tables \ref{table5}--\ref{table7} summarize results
for PennCNV, fused-lasso minimization and DPI, respectively. As with
data set 1, PennCNV achieves the best sensitivity, followed by DPI. The
best control of false positives occurs with DPI. The accuracy of the
methods and the optimal values of $\lambda_1$ and $\lambda_2$ do not
change with sequence length~$n$. However, it is clear that the
advantages of selecting individual-specific $\lambda$ values outweigh
the benefit of selecting constant $\lambda$ values that maximize
overall performance.  In fact, the choice of the oracle $\lambda$ is
excessively influenced by some individuals with poor quality data; to
control false discoveries in these subjects, one lowers performance in
more favorable settings.
\begin{table}\tablewidth=208pt
\caption{Accuracy of PennCNV for various SNP sequence lengths}\label{table5}
\begin{tabular*}{208pt}{@{\extracolsep{\fill}}lccc@{}}
\hline
\textbf{Sequence length} & \textbf{TPR (\%)} & \textbf{FPR (\%)}  & \textbf{FDR (\%)} \\
\hline
\phantom{0,}4000 & 95.54 & 0.0029 & 0.46 \\
\phantom{0,}8000 & 95.43 & 0.0019 & 0.62 \\
12,000 & 96.71 & 0.0038 & 1.77 \\
16,000 & 96.46 & 0.0012 & 0.74 \\
20,000 & 95.60 & 0.0007 & 0.59 \\
Overall & 95.95 & 0.0018 & 0.84 \\
\hline
\end{tabular*}
\end{table}

\begin{table}[b]
\tabcolsep=0pt
\caption{Accuracy of fused-lasso minimization for various SNP sequence
lengths. For strategy \textup{(a)}, average values of $\lambda_1$ and
$\lambda_2$ specified for each individual are summarized for each SNP
sequence length; Standard errors appear in parentheses} \label{table6}
\begin{tabular*}{\textwidth}{@{\extracolsep{\fill}}lccccc@{}}
\hline
\textbf{Sequence length} & $\bolds{\lambda_1}$ & $\bolds{\lambda_2}$ & \textbf{TPR (\%)} & \textbf{FPR (\%)}  & \textbf{FDR (\%)} \\
\hline
\multicolumn{6}{@{}l}{(a) $\lambda1$ and $\lambda2$ specified for each individual according to equation (\ref{L1L2choice})} \\
  \phantom{0,}4000 & 0.13 (0.04) & 0.73 (0.23) & 88.40 & 0.0414 & 6.73 \\
  \phantom{0,}8000 & 0.13 (0.04) & 0.76 (0.24) & 89.54 & 0.0241 & 7.66 \\
12,000 & 0.12 (0.03) & 0.76 (0.16) & 90.85 & 0.0148 & 7.00 \\
16,000 & 0.13 (0.04) & 0.79 (0.22) & 87.63 & 0.0103 & 6.77 \\
20,000 & 0.13 (0.04) & 0.80 (0.22) & 85.34 & 0.0084 & 7.07 \\
Overall & -- & -- & 88.35 & 0.0145 & 7.05 \\ [5pt]
\multicolumn{6}{@{}l}{(b) Oracle choice of $\lambda1$ and $\lambda2$: constant values across all individuals} \\
 \phantom{0,}4000 & 0.16 & 0.80 & 83.70 & 0.0414 & 7.08 \\
  \phantom{0,}8000 & 0.19 & 0.80 & 77.46 & 0.0206 & 7.58 \\
12,000 & 0.18 & 0.80 & 84.09 & 0.0141 & 7.20 \\
16,000 & 0.17 & 0.90 & 81.12 & 0.0102 & 7.20 \\
20,000 & 0.18 & 0.80 & 76.12 & 0.0077 & 7.26 \\
Overall & -- & -- & 80.50 & 0.0136 & 7.26 \\
\hline
\end{tabular*}
\end{table}

\begin{table}
\caption{Accuracy of DPI for various SNP sequence lengths. For strategy
\textup{(a)}, average values of $\lambda_1$ and $\lambda_2$ specified for each
individual are summarized for each SNP sequence length;\break Standard errors
appear in parentheses} \label{table7}
\begin{tabular*}{\textwidth}{@{\extracolsep{\fill}}lccccc@{}}
\hline
\textbf{Sequence length} & $\bolds{\lambda_1}$ & $\bolds{\lambda_2}$ & \textbf{TPR (\%)} & \textbf{FPR (\%)}  & \textbf{FDR (\%)} \\
\hline
\multicolumn{6}{@{}l}{(a) $\lambda1$ and $\lambda2$ specified for each individual according to equation (\ref{L1L2choice})} \\
 \phantom{0,}4000 & 0.13 (0.04) & 0.73 (0.23) & 93.70 & 0.0013 & 0.22 \\
 \phantom{0,}8000 & 0.13 (0.04) & 0.76 (0.24) & 93.33 & 0.0007 & 0.22 \\
12,000 & 0.12 (0.03) & 0.76 (0.16) & 95.78 & 0.0004 & 0.22 \\
16,000 & 0.13 (0.04) & 0.79 (0.22) & 94.77 & 0.0009 & 0.56 \\
20,000 & 0.13 (0.04) & 0.80 (0.22) & 92.32 & 0.0005 & 0.43 \\
Overall & -- & -- & 93.98 & 0.0007 & 0.33 \\ [5pt]
\multicolumn{6}{@{}l}{(b) Oracle choice of $\lambda1$ and $\lambda2$: constant values across all individuals} \\
 \phantom{0,}4000 & 0.15 & 2.50 & 87.72 & 0.0013 & 0.22 \\
  \phantom{0,}8000 & 0.24 & 2.70 & 86.35 & 0.0007 & 0.25 \\
12,000 & 0.12 & 1.80 & 94.43 & 0.0004 & 0.22 \\
16,000 & 0.18 & 2.10 & 91.51 & 0.0009 & 0.60 \\
20,000 & 0.16 & 2.00 & 90.18 & 0.0005 & 0.41 \\
Overall & -- & -- & 90.04 & 0.0007 & 0.34 \\
\hline
\end{tabular*}
\end{table}

\subsection{Speed comparison of different methods for CNV detection}

\begin{figure}[b]

\includegraphics{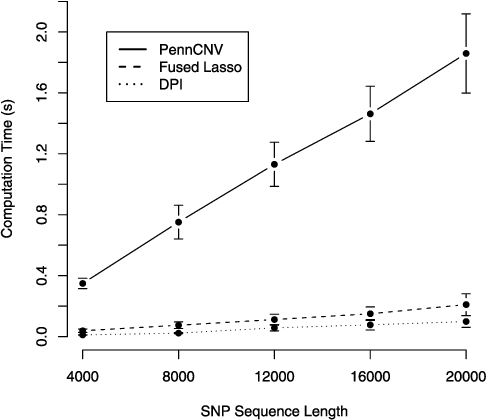}

\caption{Graphical comparison of computation speed as sequence length
varies. Solid line: PennCNV; Dashed line: Fused Lasso; Dotted line:
DPI.}\label{fig4}
\end{figure}

Finally, we compared the computational speeds of the three methods.
Although the cost of each scales linearly with the number of SNPs, run
times vary considerably in practice (see Figure \ref{fig4}). We base
our comparisons on data set 2 run on an Intel Xeon 2.80 GHz processor
operating under Linux. The PennCNV distributed software (2008, November
19 version) is a combination of C and Perl.  We implemented DPI and the
MMTDM algorithm for fused-lasso minimization in Fortran 95. The penalty
tuning parameters were chosen according to equation (\ref{L1L2choice}).
For DPI we set $\alpha=12$. Table \ref{table8} lists average run times
for each sequence sample; standard errors appear in parentheses.  As we
anticipated, fused-lasso minimization and DPI require less computation
per iteration and run much faster than PennCNV. DPI is 2 to 3 times
faster than fused-lasso minimization.

\begin{table}
\tabcolsep=0pt
\caption{Computation times for the three CNV imputation methods. The
tuning constants in the fused lasso and DPI are noted in Section
\protect\ref{accuracy_by_length} } \label{table8}
\begin{tabular*}{\textwidth}{@{\extracolsep{\fill}}lccc@{}}
\hline
\textbf{Sequence length} & \textbf{PennCNV (s)} & \textbf{Fused Lasso (s)} & \textbf{DPI (s)} \\
\hline
  \phantom{0,}4000 & 0.349 (0.034) & 0.038 (0.011) & 0.011 (0.002) \\
  \phantom{0,}8000 & 0.751 (0.111) & 0.075 (0.022) & 0.023 (0.003) \\
12,000 & 1.131 (0.145) & 0.112 (0.035) & 0.057 (0.020) \\
16,000 & 1.462 (0.181) & 0.150 (0.045) & 0.077 (0.034) \\
20,000 & 1.859 (0.260) & 0.210 (0.072) & 0.099 (0.038) \\
\hline
\end{tabular*}
\end{table}

\subsection{Analysis of four real samples}

We tested the three methods on genome scan data on four schizophrenia
patients from the study of \citet{schizoAJHG08}. These patients were
selected because they each exhibit one experimentally validated CNV
(two deletions and two duplications). The four CNVs disrupt the genes
MYT1L, CTNND2, NRXN1 and ASTN2, which play important roles in neuronal
functioning and are associated with schizophrenia. This subset of the
data is ideal for our purpose. The entire data set was collected as
part of a genome-wide association study and consists of blood samples
from unrelated individuals. It is expected that only a modest amount of
CNV may be present; most CNVs probably represent inherited neutral
polymorphisms rather de novo mutations. Unlike cancer cell lines, copy
numbers should rarely exceed 3.

We analyzed the entire genomes of these four subjects, applying the
three methods to each chromosome arm. In calling CNVs with fused-lasso
minimization, we controlled FDR at the 0.05 level. The penalty tuning
parameters were chosen according to equation (\ref{L1L2choice}). For
dynamic programming, we set $\alpha=12$. It took on average $113.8$,
$8.6$ and $4.7$ seconds for the three methods to run on the
approximately 550k SNPs typed on each individual.  The computational
efficiency of DPI displayed here may be a decisive advantage in other
data sets with thousands of participants. To focus on signals with a
higher chance of being real, we eliminated all CNV calls involving
fewer than $5$ SNPs.

Table \ref{table9} reports the numbers of detected CNVs and their
median sizes; median absolute deviations are listed in parentheses.
PennCNV produced the largest number of CNVs calls, followed by
fused-lasso minimization. The CNVs detected by PennCNV and DPI had
similar sizes; those detected by fused-lasso minimization tended to be
longer. Table \ref{table10} summarizes the overlap between the CNVs
calls for the three methods. The vast majority of CNVs detected by DPI
are also detected by PennCNV. There is a smaller overlap between
PennCNV and the Fused Lasso.

\begin{figure}
\tabcolsep=0pt
\begin{tabular*}{\textwidth}{@{\extracolsep{\fill}}cc@{}}
(a)& (b)\\ [-1 pt]

\includegraphics{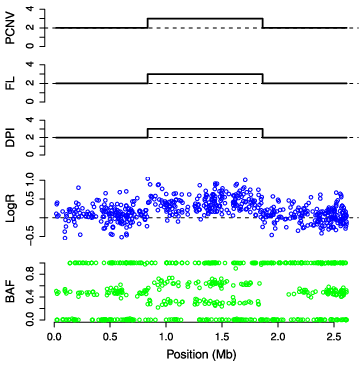}
&\includegraphics{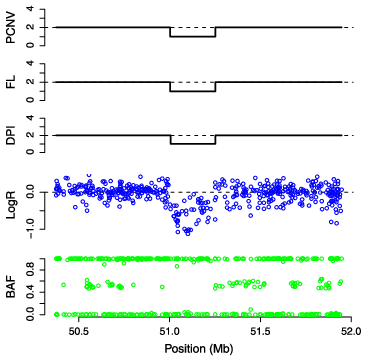}\\ [5pt]
(c)& (d)\\ [-1 pt]

\includegraphics{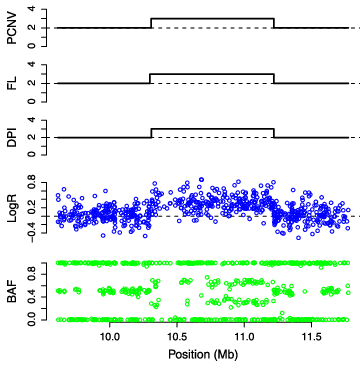}
&\includegraphics{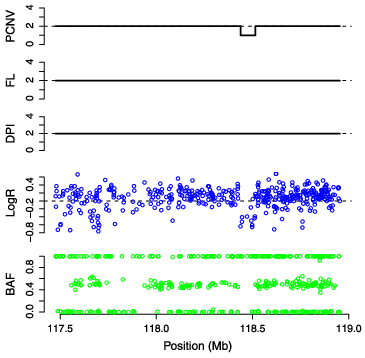}\\
\end{tabular*}
\caption{PennCNV, fused-lasso minimization, and DPI detected
experimentally verified CNVs in 4 schizophrenia patients: \textup{(a)} A
duplication on 2p25.3 of Patient 1; \textup{(b)} A deletion on 2p16.3 of Patient~2;
\textup{(c)} A duplication on 5p15.2 of Patient 3; \textup{(d)} A deletion on 9q33.1
of Patient 4. In each subplot from top to bottom, the first three
panels display the CNV detected by PennCNV, fused-lasso minimization
and DPI respectively, the fourth panel displays in blue $y_i$ (LogR),
and the fifth panel displays in green $x_i$ (BAF).}\label{real}
\vspace*{-2pt}
\end{figure}

Three of the experimentally verified CNVs were detected by all three
methods.  The fourth, a deletion on 9q33.1 in patient 4, was detected
only by PennCNV (see Figure \ref{real}). It is noteworthy that the
quality of the data for this patient is poor. For example, it fails to
pass the PennCNV quality control criterion requiring the standard
deviation of LogR to be less than 0.2. In this sample the standard
deviation is 0.26.  It appears that the higher sensitivity of PennCNV
comes at the price of allowing too many false positives. PennCNV calls
an exceptionally high number (85) of CNVs for patient 4, with limited
overlap with the other two methods.

\begin{table}
\tabcolsep=0pt
\caption{CNVs detected by PennCNV, Fused Lasso and DPI for each
patient} \label{table9}
\begin{tabular*}{\textwidth}{@{\extracolsep{\fill}}lcccccc@{}}
\hline
& \multicolumn{2}{c}{\textbf{PennCNV}} & \multicolumn{2}{c}{\textbf{Fused Lasso}} & \multicolumn{2}{c}{\textbf{DPI}}
\\ [-7 pt]
&\multicolumn{2}{@{}c}{\hrulefill}&\multicolumn{2}{c}{\hrulefill}&\multicolumn{2}{c@{}}{\hrulefill}\\
\textbf{Patient} & \textbf{\#CNV} & \textbf{CNV size} & \textbf{\#CNV} & \textbf{CNV size}  & \textbf{\#CNV} & \textbf{CNV size} \\
\hline
1 & 34 & 8 (4) & 18 & 17 (7)   & 16 & 10 (4) \\
2 & 12 & 7 (3) & 13 & 11 (9)   &  \phantom{0}7 & \phantom{0}7 (3) \\
3 & 19 & 8 (4) & 14 & \phantom{0}18 (16) &  22 & \phantom{0}7 (2) \\
4 & 85 & 8 (4) & 20 & \phantom{0}19 (16) &  18 & \phantom{0}9 (4) \\
\hline
\end{tabular*}
\end{table}

\begin{table}[b]
\caption{Overlap of CNVs detected by PennCNV, Fused Lasso and DPI. The
percentages listed in parentheses refer to the ratio of the number of
overlapping CNVs to the total number of unique CNVs detected. For
patient 1 DPI treated a large duplication region on the long arm of
Chromosome 22 as two segments. Thus, the number of overlapping CNVs was\break
increased by $1$ compared to PennCNV vs Fused Lasso} \label{table10}
\begin{tabular*}{\textwidth}{@{\extracolsep{\fill}}lccccc@{}}
\hline
\textbf{Patient} & \textbf{PennCNV} & \textbf{PennCNV} & \textbf{Fused Lasso} & \textbf{3 methods} \\
 &  \textbf{vs Fused Lasso} & \textbf{vs DPI} &  \textbf{vs DPI} & \\
\hline
1 &   \phantom{0}7 (15.6\%) & 12 (31.6\%) &   9 (36.0\%) &   \phantom{0}8 (16.7\%) \\
2 &   \phantom{0}7 (38.9\%) &   \phantom{0}6  (46.2\%) &   7 (53.8\%) &   \phantom{0}6 (33.3\%) \\
3 & 10 (43.5\%) & 15 (57.7\%) &   8 (28.6\%) &   \phantom{0}8 (26.7\%) \\
4 &   8 (8.2\%)    & 13 (14.4\%) &   8 (26.7\%) &   7 (6.9\%) \\
\hline
\end{tabular*}
\end{table}

\section{Discussion} \label{discussion}

We have proposed two new methods for the reconstruction of CNV. Both
methods are much faster than PennCNV, the current state-of-the-art
method in CNV discovery. The greater accuracy of DPI versus fused-lasso
minimization underscores the importance of using BAF measurements and
capitalizing on the discrete nature of CNV imputation. DPI has the
additional advantage of outputting the allelic copy numbers so helpful
in refining the associations between CNVs and phenotypes.  It is hardly
surprising that DPI exhibits superior performance in the schizophrenia
data where its underlying assumptions hold. By contrast in the analysis
of tumor cells, it is much more difficult to fix a priori the number of
copies. With its flexibility in fitting piecewise constant functions to
LogR intensities, the fused lasso will shine in this less discrete
setting.

We would like to emphasize that both proposed methods are rough
compared to well-established algorithms like PennCNV. There is
definitely room for further performance improvements by redefining the
loss and penalty functions. As a concrete example, one could modify the
fused-lasso penalties to reflect the distances between adjacent SNPs
[\citet{fqrCGH}]. We suggest scaling the difference
$|\beta_i-\beta_{i-1}|$ by the reciprocal of the physical distance
$|b_i-b_{i-1}|$.  Anyone wanting to use or modify our statistical
procedures is welcome to our Fortran source code. Please contact the
first author for a copy.

We can expect to see more applications of penalized estimation
throughout genomics. In our view, penalized models are more
parsimonious than hidden Markov models and achieve many of the same
aims. Our redefinition of the fused-lasso penalty and application of
the MM algorithm circumvent some of the toughest issues of penalized
estimation in the CNV context and have important implications for other
application areas such as time series analysis. For more traditional
theoretical and numerical approaches to penalized estimation, we
recommend the recent survey paper on $\ell_1$ trend filtering
[\citet{kim09}].

\section*{Acknowledgments}
We thank Joe de Young for help in reclustering the genotype signal and
Jacobine Buizer-Voskamp for providing us with detailed information in
real data analysis. We acknowledge the editor and the anonymous
reviewer for their constructive comments on this manuscript.


\printaddresses


\begin{thebibliography}{26}

\bibitem[\protect\citeauthoryear{Bioucas-Diaa, Figueiredo and
  Oliveira}{2006}]{bioucas06}
  \textsc{Bioucas-Diaa}, J. M.,
  \textsc{Figueiredo}, M. A. T. and
  \textsc{Oliveira}, J. P.
(2006). Adaptive total variation image deconvolution: A
  majorization--minimization approach.
In \textit{IEEE International Conference on Acoustics, Speech, and
Signal Processing (ICASSP'06)}. Toulouse, France.



\bibitem[\protect\citeauthoryear{Cand\`{e}s and Plan}{2009}]{Candes}
\textsc{Cand\`{e}s}, E. J. and
  \textsc{Plan}, Y.
(2009). Near-ideal model selection by $\ell_1$ minimization.
\textit{Ann. Statist.} \textbf{37} 2145--2177.
\MR{2543688}



\bibitem[\protect\citeauthoryear{Chan and Shen}{2002}]{inpainting}
  \textsc{Chan}, T. F. and
  \textsc{Shen}, J.
(2002). Mathematical models for local nontexture inpainting.
\textit{SIAM J. Appl. Math.} \textbf{62} 1019--1043.
\MR{1897733}


\bibitem[\protect\citeauthoryear{Colella  et al.}{2007}]{colella07}
  \textsc{Colella}, S.,
  \textsc{Yau}, C.,
  \textsc{Taylor}, J. M.,
  \textsc{Mirza}, G.,
  \textsc{Butler}, H.,
  \textsc{Clouston}, P.,
  \textsc{Bassett}, A.~S.,
  \textsc{Seller}, A.,
  \textsc{Holmes}, C. C. and
  \textsc{Ragoussis}, J.
(2007).
QuantiSNP: An objective Bayes hidden-Markov model to detect and
  accurately map copy number variation using SNP genotyping data.
\textit{Nucleic Acids Research} \textbf{35} 2013--2025.



\bibitem[\protect\citeauthoryear{Conte and deBoor}{1972}]{TDMA}
\textsc{Conte}, S. D. and
\textsc{deBoor}, C. (1972).
\textit{Elementary Numerical Analysis}. McGraw-Hill, New York.



\bibitem[\protect\citeauthoryear{Diskin  et al.}{2008}]{gcadj}
  \textsc{Diskin}, S. J.,
  \textsc{Li}, M.,
  \textsc{Hou}, C.,
  \textsc{Yang}, S.,
  \textsc{Glessner}, J.,
  \textsc{Hakonarson}, H.,
  \textsc{Bucan}, M.,
  \textsc{Maris}, J. M. and
  \textsc{Wang}, K.
(2008). Adjustment of genomic waves in signal intensities from
whole-genome SNP
  genotyping platforms.
\textit{Nucleic Acids Research} \textbf{36} e126.



\bibitem[\protect\citeauthoryear{Donoho and Johnstone}{1994}]{Donoho}
\textsc{Donoho}, D. L. and
  \textsc{Johnstone}, I. M.
(1994). Ideal spatial adaptation by wavelet shrinkage.
\textit{Biometrika} \textbf{81} 425--455.
\MR{1311089}



\bibitem[\protect\citeauthoryear{Friedman  et al.}{2007}]{pco}
\textsc{Friedman}, J.,
  \textsc{Hastie}, T.,
  \textsc{H\"{o}fling}, H. and
  \textsc{Tibshirani}, R.
(2007). Pathwise coordinate optimization. \textit{Ann. Appl.
Statist.} \textbf{1} 302--332.
\MR{2415737}



\bibitem[\protect\citeauthoryear{Iafrate  et al.}{2004}]{Iafrate04}
\textsc{Iafrate}, A. J.,
  \textsc{Feuk}, L.,
  \textsc{Rivera}, M. N.,
  \textsc{Listewnik}, M. L.,
  \textsc{Donahoe}, P. K.,
  \textsc{Qi}, Y.,
  \textsc{Scherer}, S. and
  \textsc{Lee}, C.
(2004). Detection of large-scale variation in the human genome.
\textit{Nature Genetics} \textbf{36} 949--951.



\bibitem[\protect\citeauthoryear{Jakobsson  et al.}{2008}]{HapMapCNV}
  \textsc{Jakobsson}, M.,
  \textsc{Scholz}, S. W.,
  \textsc{Scheet}, P.,
  \textsc{Gibbs}, J. R.,
  \textsc{VanLiere}, J. M.,
  \textsc{Fung}, H. C.,
  \textsc{Szpiech}, Z. A.,
  \textsc{Degnan}, J. H.,
  \textsc{Wang}, K.,
  \textsc{Guerreiro}, R.,
  \textsc{Bras}, J. M.,
  \textsc{Schymick}, J. C.,
  \textsc{Hernandez}, D. G.,
  \textsc{Traynor}, B. J.,
  \textsc{Simon-Sanchez}, J.,
  \textsc{Matarin}, M.,
  \textsc{Britton}, A.,
  \textsc{van de Leemput}, J.,
  \textsc{Rafferty}, I.,
  \textsc{Bucan}, M.,
  \textsc{Cann}, H. M.,
  \textsc{Hardy}, J. A.,
  \textsc{Rosenberg}, N. A. and
  \textsc{Singleton}, A. B.
(2008). Genotype, haplotype and copy-number variation in worldwide
human populations.
\textit{Nature} \textbf{451} 998--1003.



\bibitem[\protect\citeauthoryear{Kim  et al.}{2009}]{kim09}
  \textsc{Kim}, S.-J.,
  \textsc{Koh}, K.,
  \textsc{Boyd}, S. and
  \textsc{Gorinevsky}, D.
(2009). $\ell_1$ trend filtering. \textit{SIAM Review} \textbf{51}
339--360.
\MR{2505584}



\bibitem[\protect\citeauthoryear{Korn  et al.}{2008}]{bird}
  \textsc{Korn}, J. M.,
  \textsc{Kuruvilla}, F. G.,
  \textsc{McCarroll}, S. A.,
  \textsc{Wysoker}, A.,
  \textsc{Nemesh}, J.,
  \textsc{Cawley}, S.,
  \textsc{Hubbell}, E.,
  \textsc{Veitch}, J.,
  \textsc{Collins}, P. J.,
  \textsc{Darvishi}, K.,
  \textsc{Lee}, C.,
  \textsc{Nizzari}, M. M.,
  \textsc{Gabriel}, S. B.,
  \textsc{Purcell}, S.,
  \textsc{Daly}, M. J. and
  \textsc{Altshuler}, D. D.
(2008). Integrated genotype calling and association analysis of SNPs,
common
  copy number polymorphisms and rare CNVs.
\textit{Nature Genetics} \textbf{40} 1253--1260.



\bibitem[\protect\citeauthoryear{Lange}{2004}]{LangeOpt}
\textsc{Lange}, K. (2004). \textit{Optimization}. Springer, New York.
\MR{2072899}


\bibitem[\protect\citeauthoryear{Li and Zhu}{2007}]{fqrCGH}
  \textsc{Li}, Y. and
  \textsc{Zhu}, J.
(2007). Analysis of array CGH data for cancer studies using fused
quantile
  regression.
\textit{Bioinformatics} \textbf{23} 2470--2476.



\bibitem[\protect\citeauthoryear{Negahban  et al.}{2009}]{Wainwright}
  \textsc{Negahban}, S.,
  \textsc{Ravikmuar}, P.,
  \textsc{Wainwright}, M. J. and
  \textsc{Yu}, B.
(2009). A unified framework for high-dimensional analysis of
M-estimators with
  decomposable regularizers.
In \textit{The Neural Information Processing Systems Conference
(NIPS'09)}. Vancouver, Canada.



\bibitem[\protect\citeauthoryear{Redon  et al.}{2006}]{redon06}
  \textsc{Redon}, R.,
  \textsc{Ishikawa}, S.,
  \textsc{Fitch}, K. R.,
  \textsc{Feuk}, L.,
  \textsc{Perry}, G. H.,
  \textsc{Andrews}, T. D.,
  \textsc{Fiegler}, H.,
  \textsc{Shapero}, M. H.,
  \textsc{Carson}, A. R.,
  \textsc{Chen}, W.,
  \textsc{Cho}, E. K.,
  \textsc{Dallaire}, S.,
  \textsc{Freeman}, J. L.,
  \textsc{Gonzalez}, J. R.,
  \textsc{Gratacos}, M.,
  \textsc{Huang}, J.,
  \textsc{Kalaitzopoulos}, D.,
  \textsc{Komura}, D.,
  \textsc{MacDonald}, J. R.,
  \textsc{Marshall}, C. R.,
  \textsc{Mei}, R.,
  \textsc{Montgomery}, L.,
  \textsc{Nishimura}, K.,
  \textsc{Okamura}, K.,
  \textsc{Shen}, F.,
  \textsc{Somerville},  M. J.,
  \textsc{Tchinda}, J.,
  \textsc{Valsesia}, A.,
  \textsc{Woodwark}, C.,
  \textsc{Yang}, F.,
  \textsc{Zhang}, J.,
  \textsc{Zerjal}, T.,
  \textsc{Zhang}, J.,
  \textsc{Armengol}, L.,
  \textsc{Conrad}, D. F.,
  \textsc{Estivill}, X.,
  \textsc{Tyler-Smith}, C.,
  \textsc{Carter}, N. P.,
  \textsc{Aburatani}, H.,
  \textsc{Lee}, C.,
  \textsc{Jones}, K. W.,
  \textsc{Scherer}, S. W. and
  \textsc{Hurles}, M. E.
(2006). Global variation in copy number in the human genome.
\textit{Nature} \textbf{444} 444--454.



\bibitem[\protect\citeauthoryear{Rudin, Osher and Fatemi}{1992}]{rof}
  \textsc{Rudin}, L. I.,
  \textsc{Osher}, S. and
  \textsc{Fatemi}, E.
(1992). Nonlinear total variation based noise removal algorithms.
\textit{Physica D} \textbf{60} 259--268.



\bibitem[\protect\citeauthoryear{Scharpf  et al.}{2008}]{ingo}
  \textsc{Scharpf}, R. B.,
  \textsc{Parmigiani}, G.,
  \textsc{Pevsner}, J. and
  \textsc{Ruczinski}, I.
(2008). Hidden Markov models for the assessment of chromosomal
alterations
  using high throughput SNP arrays.
\textit{Ann. Appl. Statist.} \textbf{2} 687--713.
\MR{2524352}



\bibitem[\protect\citeauthoryear{Sebat  et al.}{2004}]{sebat04}
  \textsc{Sebat}, J.,
  \textsc{Lakshmi}, B.,
  \textsc{Troge}, J.,
  \textsc{Alexander}, J.,
  \textsc{Young}, J.,
  \textsc{Lundin}, P.,
  \textsc{Maner}, S.,
  \textsc{Massa}, H.,
  \textsc{Walker}, M.,
  \textsc{Chi}, M.,
  \textsc{Navin}, N.,
  \textsc{Lucito}, R.,
  \textsc{Healy}, J.,
  \textsc{Hicks}, J.,
  \textsc{Ye}, K.,
  \textsc{Reiner}, A.,
  \textsc{Gilliam}, T. C.,
  \textsc{Trask}, B.,
  \textsc{Patterson}, N.,
  \textsc{Zetterberg}, A. and
  \textsc{Wigler}, M.
(2004). Large-scale copy number polymorphism in the human genome.
\textit{Science} \textbf{305} 525--528.



\bibitem[\protect\citeauthoryear{Stefansson  et al.}{2008}]{nature}
  \textsc{Stefansson}, H.,
  \textsc{Rujescu}, D.,
  \textsc{Cichon}, S.,
  \textsc{Pietil\"{a}inen}, O. P. H.,
  \textsc{Ingason}, A.,
  \textsc{Steinberg}, S.,
  \textsc{Fossdal}, R.,
  \textsc{Sigurdsson}, E.,
  \textsc{Sigmundsson}, T.,
  \textsc{Buizer-Voskamp}, J. E.,
  \textsc{Hansen}, T.,
  \textsc{Jakobsen}, K. D.,
  \textsc{Muglia}, P.,
  \textsc{Francks}, C.,
  \textsc{Matthews}, P. M.,
  \textsc{Gylfason}, A.,
  \textsc{Halldorsson}, B. V.,
  \textsc{Gudbjartsson}, D.,
  \textsc{Thorgeirsson}, T. E.,
  \textsc{Sigurdsson}, A.,
  \textsc{Jonasdottir}, A.,
  \textsc{Jonasdottir}, A.,
  \textsc{Bjornsson}, A.,
  \textsc{Mattiasdottir}, S.,
  \textsc{Blondal}, T.,
  \textsc{Haraldsson}, M.,
  \textsc{Magnusdottir}, B. B.,
  \textsc{Giegling}, I.,
  \textsc{M\"{o}ller}, H.-J.,
  \textsc{Hartmann}, A.,
  \textsc{Shianna}, K. V.,
  \textsc{Ge}, D.,
  \textsc{Need}, A. C.,
  \textsc{Crombie}, C.,
  \textsc{Fraser}, G.,
  \textsc{Walker}, N.,
  \textsc{Lonnqvist}, J.,
  \textsc{Suvisaari}, J.,
  \textsc{Tuulio-Henriksson}, A.,
  \textsc{Paunio}, T.,
  \textsc{Toulopoulou}, T.,
  \textsc{Bramon}, E.,
  \textsc{Di Forti}, M.,
  \textsc{Murray}, R.,
  \textsc{Ruggeri}, M.,
  \textsc{Vassos}, E.,
  \textsc{Tosato}, S.,
  \textsc{Walshe}, M.,
  \textsc{Li}, T.,
  \textsc{Vasilescu}, C.,
  \textsc{M\"{u}hleisen}, T. W.,
  \textsc{Wang}, A. G.,
  \textsc{Ullum}, H.,
  \textsc{Djurovic}, S.,
  \textsc{Melle}, I.,
  \textsc{Olesen}, J.,
  \textsc{Kiemeney}, L. A.,
  \textsc{Franke}, B.,
  \textsc{Genetic Risk
  and Outcome in Psychosis (GROUP)},
  \textsc{Sabatti}, C.,
  \textsc{Freimer}, N. B.,
  \textsc{Gulcher}, J. R.,
  \textsc{Thorsteinsdottir}, U.,
  \textsc{Kong}, A.,
  \textsc{Andreassen}, O. A.,
  \textsc{Ophoff}, R. A.,
  \textsc{Georgi}, A.,
  \textsc{Rietschel}, M.,
  \textsc{Werge}, T.,
  \textsc{Petursson}, H.,
  \textsc{Goldstein}, D. B.,
  \textsc{N\"{o}then}, M. M.,
  \textsc{Peltonen}, L.,
  \textsc{Collier}, D. A.,
  \textsc{St Clair}, D. and
  \textsc{Stefansson}, K.
(2008). Large recurrent microdeletions associated with schizophrenia.
\textit{Nature} \textbf{455} 232--236.



\bibitem[\protect\citeauthoryear{Tibshirani and Wang}{2008}]{cghFLasso}
  \textsc{Tibshirani}, R. and
  \textsc{Wang}, P.
(2008). Spatial smoothing and hot spot detection for CGH data using the
Fused
  Lasso.
\textit{Biostatistics} \textbf{9} 18--29.



\bibitem[\protect\citeauthoryear{Tibshirani  et al.}{2005}]{FLasso}
  \textsc{Tibshirani}, R.,
  \textsc{Saunders}, M.,
  \textsc{Rosset}, S.,
  \textsc{Zhu}, J. and
  \textsc{Knight}, K.
(2005). Sparsity and smoothness via the fused lasso. \textit{J.
Roy. Statist. Soc. Ser. B} \textbf{67} 91--108.
\MR{2136641}



\bibitem[\protect\citeauthoryear{Vrijenhoek  et al.}{2008}]{schizoAJHG08}
  \textsc{Vrijenhoek}, T.,
  \textsc{Buizer-Voskamp}, J. E.,
  \textsc{van der Stelt}, I.,
  \textsc{Strengman}, E.,
  \textsc{Genetic
  Risk and Outcome in Psychosis (GROUP) Consortium},
  \textsc{Sabatti}, C.,
  \textsc{van Kessel}, A. G.,
  \textsc{Brunner}, H. G.,
  \textsc{Ophoff}, R. A. and
  \textsc{Veltman}, J. A.
(2008). Recurrent CNVs disrupt three candidate genes in schizophrenia
  patients.
\textit{The American Journal of Human Genetics} \textbf{83} 504--510.



\bibitem[\protect\citeauthoryear{Wang  et al.}{2007}]{pennCNV}
  \textsc{Wang}, K.,
  \textsc{Li}, M.,
  \textsc{Hadley}, D.,
  \textsc{Liu}, R.,
  \textsc{Glessner}, J.,
  \textsc{Grant}, S. F. A.,
  \textsc{Hakonarson}, H. and
  \textsc{Bucan}, M.
(2007). PennCNV: An integrated hidden Markov model designed for
high-resolution
  copy number variation detection in whole-genome SNP genotyping data.
\textit{Genome Research} \textbf{17} 1665--1674.



\bibitem[\protect\citeauthoryear{Wang  et al.}{2009}]{us}
  \textsc{Wang}, H.,
  \textsc{Veldink}, J. H.,
  \textsc{Blauw}, H.,
  \textsc{van den Berg}, L. H.,
  \textsc{Ophoff}, R. A. and
  \textsc{Sabatti}, C.
(2009). Markov models for inferring copy number variations from
genotype data
  on Illumina platforms.
\textit{Human Heredity} \textbf{68} 1--22.



\bibitem[\protect\citeauthoryear{Wu and Lange}{2008}]{wulange}
  \textsc{Wu}, T. T. and
  \textsc{Lange}, K.
(2008). Coordinate descent algorithm for lasso penalized regression.
\textit{Ann. Appl. Statist.} \textbf{2} 224--244.
\MR{2415601}



\end{thebibliography}
\end{document}